# Allegro-FM: Towards Equivariant Foundation Model for Exascale Molecular Dynamics Simulations

Ken-ichi Nomura[1], Shinnosuke Hattori[2], Satoshi Ohmura[3], Ikumi Kanemasu[3], Kohei Shimamura[4], Nabankur Dasgupta[1], Aiichiro Nakano[1], Rajiv K. Kalia[1], and Priya Vashishta[1]

[1]*Collaboratory for Advanced Computing and Simulation, University of Southern California, Los Angeles, California 90089-0242, United States*

[2]*Advanced Research Laboratory, Research Platform, Sony Group Corporation, Atsugi, Kanagawa 243-0014, Japan*

[3]*Department of Civil and Environmental Engineering, Hiroshima Institute of Technology, Hiroshima 731-5193, Japan*

[4]*Department of Physics, Kumamoto University, Kumamoto 860-8555, Japan*

**Abstract**

We present a foundation model for exascale molecular dynamics simulations by leveraging an E(3) equivariant network architecture (Allegro) and a set of large-scale organic and inorganic materials datasets merged by Total Energy Alignment (TEA) framework. Thanks to the large-scale training sets, the obtained model (Allegro-FM) is versatile for various materials simulations for diverse downstream tasks covering 89 elements in the periodic table. Allegro-FM exhibits excellent agreements with high-level quantum chemistry theories in describing structural, mechanical, and thermodynamic properties, while exhibiting emergent capabilities for structural correlations, reaction kinetics, mechanical strengths, fracture, and solid/liquid dissolution, for which the model has not been trained. Furthermore, we demonstrate the robust predictability and generalizability of Allegro-FM for chemical reactions using the Transition1x that consists of 10k organic reactions and 9.6 million configurations including transition state data, as well as calcium silicate hydrates as a testbed. With its computationally efficient, strictly-local network architecture, Allegro-FM scales up to multi-billion-atom systems with a parallel efficiency of 0.964 on the exaflop/s Aurora supercomputer at Argonne Leadership Computing Facility. The approach presented in this work demonstrates the potential of the foundation model for a novel materials design and discovery based on large-scale atomistic simulations.

## Introduction

Foundation model (FM) is a paradigm shift in Artificial Intelligence (AI)[1] and has transformed our way of model training. Unlike conventional approaches that use domain-specific datasets to perform a well-targeted single task, FMs are trained with massive datasets without a specific target application in mind. A well-trained FM, also called pre-trained model, acquires the robustness and generalizability for out-of-distribution tasks, enabling diverse downstream tasks by fine-tuning with relatively small datasets. The concept of FM originates from large language models (LLMs), such as OpenAI ChatGPT[2,3], Google Gemini[4], and Meta LLaMA[5]. With the unprecedented success of LLMs, FM that is capable of performing versatile tasks with multimodal input data has gained traction in materials and chemical research[6,7].

There are two key components for a successful FM development, i.e. advanced model architecture and large-scale datasets. Transformer equipped with the attention mechanism[8] has been widely used for LLMs with hundreds of billions of learnable parameters trained on millions of tokens. Self-supervised learning is one of enabling technologies for such LLMs by eliminating the laborious and high-cost data labeling work. Fine-tuning is a form of transfer learning, in which a part of network weights in the pretrained model are adjusted by a separately prepared smaller dataset for specialized tasks.

FM has several significant advantages over conventional models developed for a narrowly defined task. Although generating a FM is extremely resource intensive, a well-trained FM may be fine-tuned with much less resources, enabling even a user with a very modest computing resource to incorporate the model in their workflow. Also, FMs are versatile and can be used as a base model for downstream tasks[1]. As such, users do not need to start new model development from scratch for a different downstream task, thereby reducing the total cost and time investment for model development and data generation.

Recently, great strides have been made in terms of both model architecture and training datasets for materials modeling and simulations. The size and complexity of training datasets together with model sizes and the number of learnable parameters have increased by several orders of magnitude. Many large-scale datasets such as Materials Project[9,10], SPICE[11], ANI, Alexandria[12], OQMD[13], AFLOW[14] are publicly available. Modern ML architectures take advantage of Graph Neural Network (GNN), multi-body expansion strategy, and the equivariant features from molecular geometry (ACE[15], MACE[16], Nequip[17], Equiformer[18,19], Orb[20], MatterGen[21]).

In this study, we present Allegro-FM, a foundation model for exascalable MD simulations, employing the linear-scaling and computationally efficient Allegro[22] architecture trained on large-scale MPtrj and SPICE datasets aligned with Total Energy Alignment (TEA) framework[23] for exascale materials applications. Allegro achieves the state-of-the-art (SOTA) accuracy and speed based on E(3) group-theoretical equivariance and local descriptors, respectively, while its excellent stability for large-scale, long-time trajectories (or fidelity scaling) can further be enhanced by sharpness-aware training as in the Allegro-Legato model[24]. TEA achieves consistent data fusion encompassing multiple computational methods through shift-scale (or affine) transformation in a metamodel space[25], which in turn is rooted in free-energy perturbation[26] and multiscale quantum-mechanics/molecular-mechanics (QM/MM) methods[27,28]. Materials simulations often require a large number of atoms to describe key features, such as dislocation, grain boundaries, distinct phases and phase boundaries. At the same time, the model should accurately describe chemical reactions. With the SOTA scalability[29] demonstrated on Perlmutter supercomputer at the National Energy Research Scientific Computing Center (NERSC), Allegro has a great potential and particularly is suitable for materials simulations that require multi million-to-billion atoms[30–32]. Large-scale materials datasets are often divided into two categories, organic molecules and inorganic crystals. They are generated with different levels of QM theories and functionals, making them incompatible for single model training. TEA framework smoothly connects the potential energy landscapes between distinct datasets, eliminating the need for expensive data regenerations.

As a test bed, we apply Allegro-FM to a tobermorite 11Å (T11) crystal, a representative system of calcium silicate hydrates and its aqueous reactions. Among silicate materials, calcium silicates are important because of the abundance in the Earth's crust as well as the primary constituent in cement. Many forms of calcium silicates and their hydrates coexist in cementitious materials, out of which T11 crystal is often used as a representative system. Tobermorites are also found in ancient Roman concretes [33].

Recently, cement also attracted attention as a carbon-storage material because of the ability to trap carbon by the mineralization process[34]. $CO_2$ mineralization naturally occurs as part of the geochemical cycle[35] and the mechanism has been extensively studied over the years. A mechanistic understanding of $CO_2$ mineralization is crucial for ensuring safe and long-term carbon storage without gas leakage. Although the carbonation of silicates is detrimental to cement by losing the mechanical strength and eventual fracture, a recent study shows alternative binder materials to mitigate the issue[36].

First-principle quantum Molecular Dynamics (QMD)[37,38] simulations provide atomistic-level insights that complement experimental observations. Density Functional Theory (DFT)[39,40] has been routinely used to access a wide range of materials properties although the high computational cost and suboptimal algorithmic scalability prohibits performing simulations spanning sufficiently long-time and large-scale to study the rich chemistry in cementitious materials. CLAYFF[41,42], a popular empirical FF, has been developed for multicomponent minerals and their hydrates. Because of the computational efficiency and scalability, CLAYFF allows incorporating complex geometries of minerals and systems and their fluid interfaces. As an empirical FF, however, CLAYFF also suffers from poor description of chemical reactions and limited transferability for different systems, therefore non-trivial efforts are required to develop an FF for new systems with robust description of materials properties. Here, a scalable FM is a promising approach to describe both quantum-mechanically accurate materials properties and chemical reactions.

**Results**

First, we have examined the structural and mechanical properties of T11 crystal shown in Fig. 1a. T11 structure consists of covalently-bonded Si-O network containing $H_2O$ molecules with the layers of calcium ions. Fig. 1b presents the equation of state of T11 using Allegro-FM. The obtained mass density at 2.44 g/cc agrees well with an experimental value of 2.46 g/cc[43]. Using Birch–Murnaghan equation of state, the bulk modulus is estimated 60 GPa by Allegro-FM, which reasonably compares to the bulk modulus of 52.7–60.8 GPa using first-principle and empirical force field calculations[44–46].

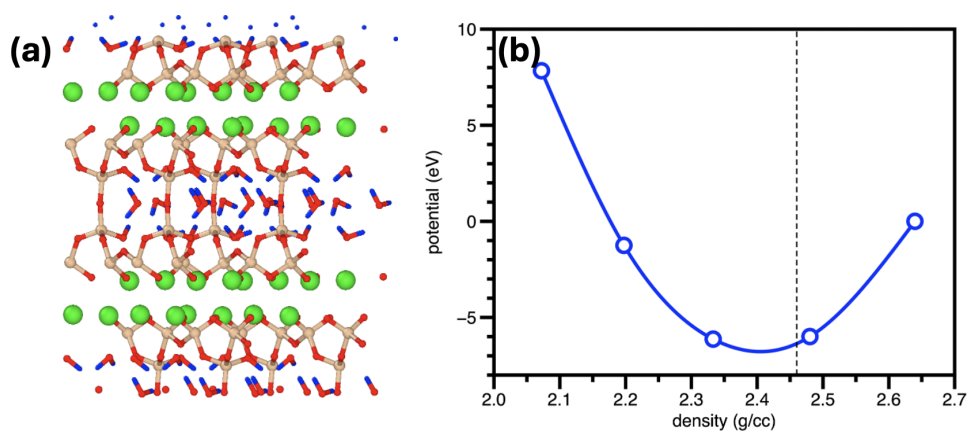

**Figure 1**. (a) A snapshot of T11 structure. Green, khaki, red, and blue spheres represent Ca, Si, O, and H atoms, respectively. (b) The equation of state using Allegro-FM. The dotted-line shows an experimental density of T11 at 2.46 g/cc[43].

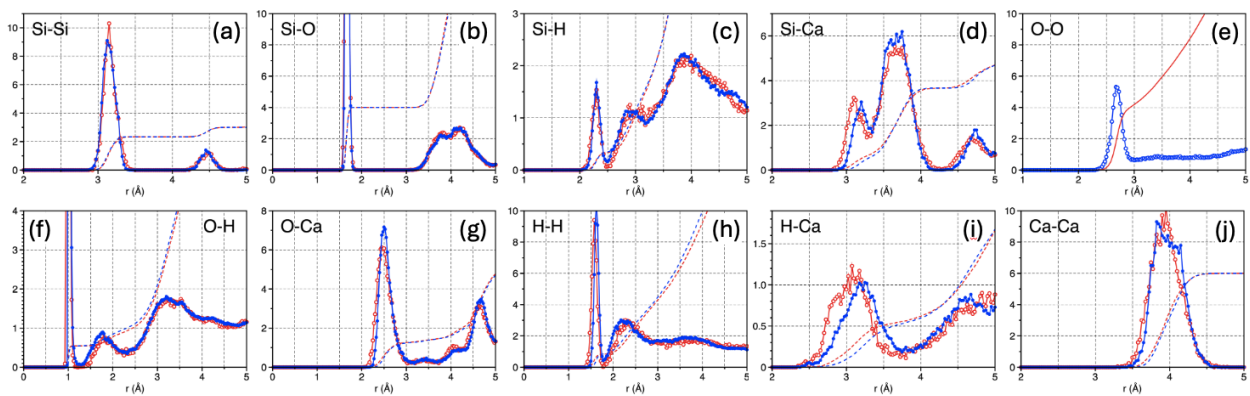

**Figure 2.** Pair distribution function (solid line) and the coordination numbers (dotted line) of T11 crystal using Allegro-FM (blue) and QMD (red). The system is thermalized at temperature 300 K with the canonical (NVT) ensemble. Beside the crystal structures indicated by the peak positions in the pair distribution functions, Allegro-FM accurately reproduces the extent of thermal motion including $H_2O$ molecules.

Figure 2 a-j, shows the pair distribution function $g(r)$ and the coordination number $N(r)$ of T11 crystal thermalized at temperature 300 K using QMD and Allegro-FM. Overall both $g(r)$ and $N(r)$ using Allegro-FM and QMD agree well. This result is somewhat expected because the thermal motion of atoms at ambient conditions would not be much different from the distribution in the training dataset. Nonetheless, the result also demonstrates the capability of Allegro-FM even without fine-tuning to accurately describe not only the crystal structure but also thermal motion of atoms at an ambient condition.

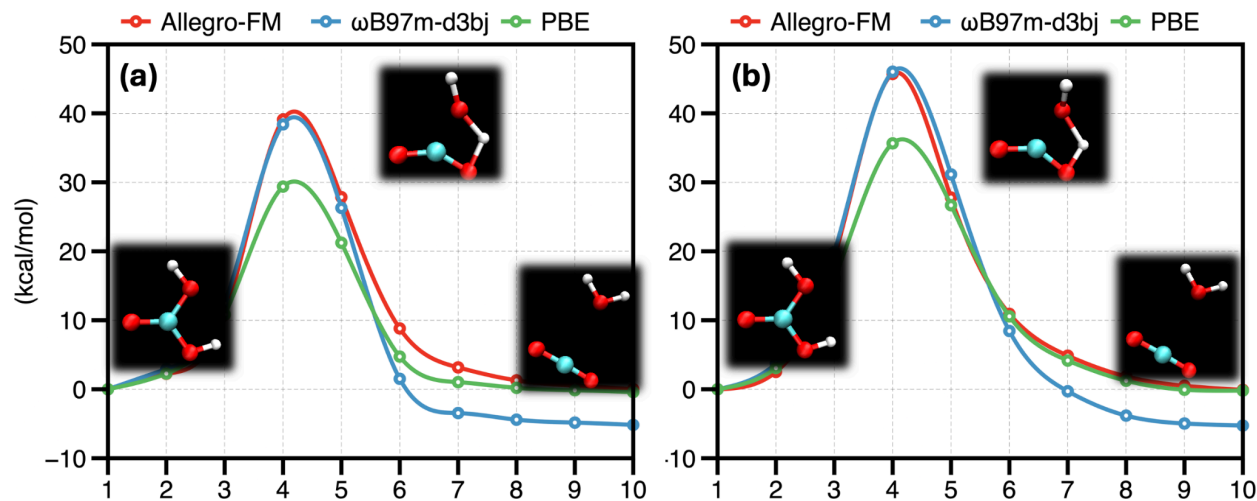

**Figure 3**. Reaction coordinate and activation energy for the decomposition reaction of a carbonic acid into $H_2O$ and $CO_2$ molecules. All results are aligned by the initial state energy as reference.

Accurate description of chemical reactions, whether a machine-learning force field (MLFF) can describe bond breaking and formation events, is a great challenge because training datasets are generally generated near ground states. Also, it is not realistic to create a training set that contains all necessary information of chemical reactions for a target system, for example, the rich multistep reactions in cement chemistry or geochemical cycles. It is a basically zero-shot learning task that requires the out-of-distribution generalizability from the training set. Here, we have examined the predictability of chemical reactions by Allegro-FM using the decomposition of carbonic acid into $H_2O$ and $CO_2$ molecules as a representative chemical reaction. Figure 3 shows the reaction coordinates obtained by the Nudged Elastic Band (NEB) method[47] using DFT. We found two distinct reaction pathways; (1) the conventional pathway that is confined within a two-dimensional plane (Fig. 3a); and (2) another pathway involving the rotation of the $H_2O$ molecule (Fig 3b). We have calculated the potential energy profile of the obtained atomic configurations using Allegro-FM, DFT with GGA-PBE and a hybrid exact-exchange functional and dispersion correlation, ωB97M-D3(BJ), functionals. Allegro-FM and ωB97M-D3(BJ) agree well on the transition state energy that may be attributed to the use of the SPICE dataset that consists of the ωB97M-D3(BJ) functional calculations for organic molecules. On the other hand, GGA-PBE and Allegro-FM agree well on the product energy, which differs by about 5 kcal/mol from the result with ωB97M-D3(BJ) functional. Overall Allegro-FM generalizes well to the transition state (TS), for example, providing the activation energy of 45.7 kcal/mol while 46.0 kcal/mol with ωB97M-D3(BJ), and 35.7 kcal/mol with GGA-PBE for the conventional pathway, respectively. More systematic analysis with a large-scale TS dataset, Transition1x[48], is provided in Supplementary Information.

In order for MLFF to be a viable method for scientific discovery, it must provide robust MD trajectory well beyond the distribution of training sets. A number of studies have reported that a high prediction accuracy does not warrant the robustness in dynamical simulations[24,49], for example, MD simulation may fail in an unpredictable manner even though an ML model shows the best benchmark result[49]. To examine the accuracy and robustness of MD trajectory predicted by Allegro-FM, we present two simulation results: (1) a tensile test on T11 crystal, and (2) a tobermorite nanoparticle (TNP) is immersed in a mixture of $H_2O$ and $CO_2$ at an elevated temperature of 1200 K.

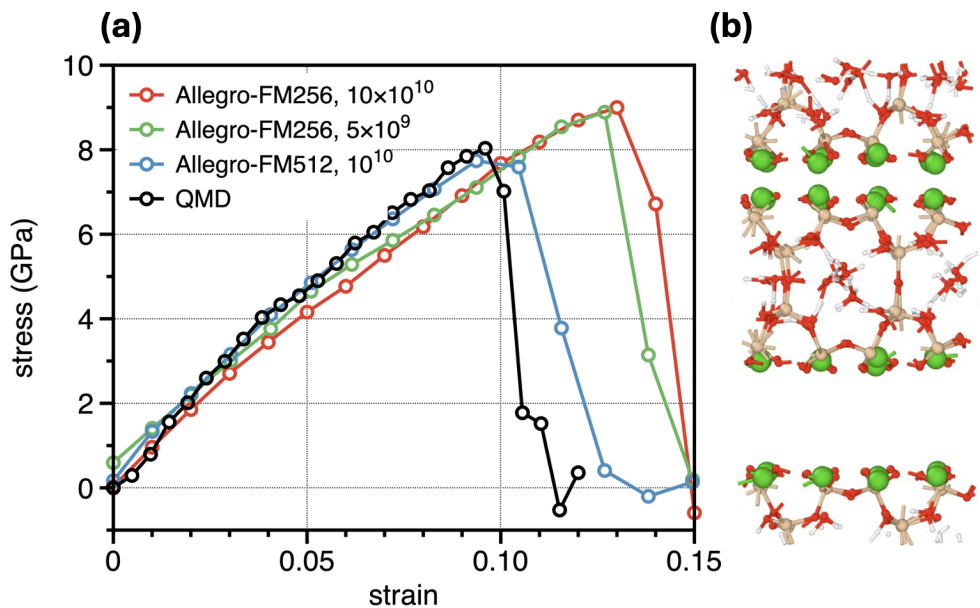

**Figure 4.** Tensile test on Tobermorite 11Å crystal thermalized at 300 K. (a) Stress-strain curves with QMD, Allegro-FM with two different sizes (256 and 512). We have examined two strain rates using Allegro-FM256 and found little effect in the stress-strain curve. The abrupt drop in stress value indicates brittle fracture. (b) A snapshot of the failed system and a clean cleavage plan at the Ca double layer.

Fracture mechanics of silicate materials have been extensively studied due to technological importance as well as its fundamental scientific questions. To examine the fracture behavior, we have performed a tensile test on a T11 crystal. The system is thermalized at 300 K and subjected to a tensile strain in the z-direction. Fig. 4a shows the stress-strain curve obtained by QMD and Allegro-FM. We have tested two model sizes (Allegro-FM256 and Allegro-FM512) and two strain rates, $10^{10}$ s$^{-1}$ and $5\times10^9$ s$^{-1}$. Overall, all systems show a clean cleavage plane between the calcium double layer. Allegro-FM and QMD show virtually identical results in the elastic regime with small strain. The strain rate effect using Allegro-FM128 is almost negligible, but the yield point is at a strain of 0.13, which is 30% larger than the DFT result. Allegro-FM512 shows an excellent agreement with the DFT result, a brittle fracture at 0.15 strain with a clean cleavage plane; see Fig. 5b.

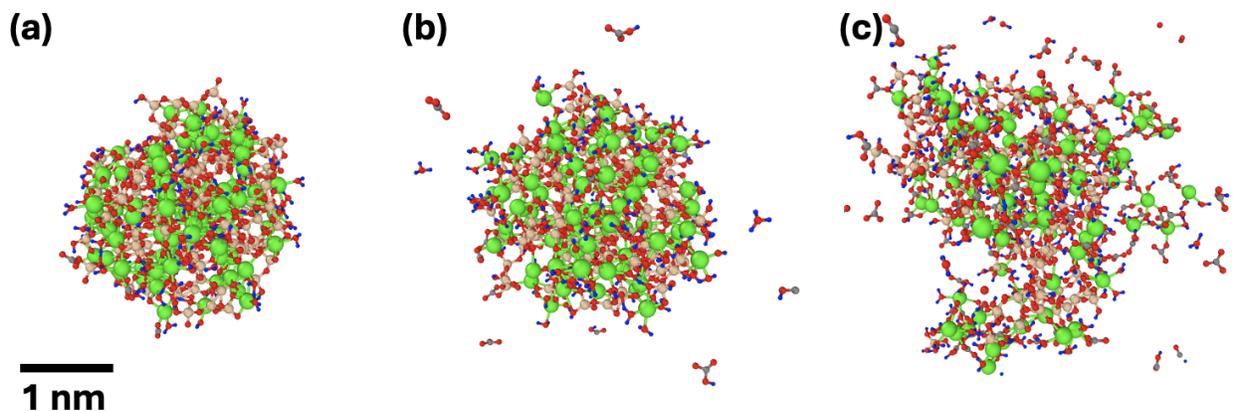

**Figure 5**. Snapshots of C-S-H nanoparticle surrounded by a mixture of $H_2O$ and $CO_2$ molecules. Ca, Si, O, and H atoms are color-coded as green, khaki, red, and blue respectively. $H_2O$ and $CO_2$ molecules are not shown for clarity. (a) the original NP shape. (b) A snapshot of NP right after the system temperature is increased to 1,200 K. An active formation of $CO_3$ and $CO_3H$ molecules around the NP is observed. (c) After 0.16 nanoseconds, many Ca atoms dissolve from the NP into the $H_2O$ and $CO_2$ mixture.

Carbonation of silicates at the solid-liquid interface involves multistep chemical reactions and is affected by many factors such as temperature, pH, and complex material geometries and surface morphology. We have carried out a surrogate simulation of the carbonation reaction process, in which a nanoparticle of T11 is immersed in a mixture of $H_2O$ and $CO_2$ molecules. The system is heated from 300 K up to 1,200 K, and then kept at 1,200 K to accelerate the overall reaction. Fig. 5a shows the initial shape of the NP. After elevating temperature to 1,200 K, we observe the formation of carbonate and bicarbonate molecules around the NP; see Fig. 5b. The leaching of Ca ions and the dissolution of the NP into the $H_2O$ and $CO_2$ has been observed after 0.16 ns shown in Fig. 5c. The obtained MD trajectory was very robust and no spurious event such as overlapping atoms nor abrupt simulation failure has been observed.

**Discussion**

Allegro-FM can perform versatile tasks beyond the original distribution of training datasets, evidenced by the simulations of activation energy, tensile test, and aqueous reactions of a calcium silicate nanoparticle. Though the training datasets do not include the information of transition states and reaction coordinates, Allegro-FM exhibits the robust predictability and generalizability for reactions that are important in the geochemical cycle and cement chemistry, demonstrating the bare capability of Allegro-FM even without fine-tuning. Unlike a conventional ML model that is trained for a narrowly-defined task, Allegro-FM is compatible with 89 elements in the periodic table. Therefore incorporating other important metals such as Al, Fe, Mg, and Ti into the aforementioned calculations does not require time-consuming data generation and expensive model training from scratch. We have also observed that the current Allegro-FM shows a suboptimal performance for magnesium silicate systems, and are currently fine-tuning it to enhance the prediction accuracy is an ongoing work.

Equipped with the robust predictability, generalizability, computational efficiency, and algorithmic scalability, Allegro-FM enables dynamical simulations spanning the micrometer spatial extent for microseconds timescale without sacrificing the atomistic details. This framework may be used to study the nanostructure of calcium silicate gel[50], reaction-induced fracture[51], self-healing cement[52], and durable cement design[33], thereby providing a novel approach for geophysical science and civil engineering applications.

**Methodology**

**- Quantum Molecular Dynamics (QMD)**

The electronic states were calculated using the projector augmented method (PAW)[53,54] within the framework of the density functional theory (DFT) in which the generalized gradient approximation (GGA-PBE)[55] was used for the exchange-correlation energy. The planewave cutoff energies for pseudo wavefunction and pseudo charge density were 30 and 200 Ry for tobermorite system and 30 and 250 Ry for $H_2O$-$CO_2$ system, respectively. In the MD simulations, the equations of motion for atoms were solved via an explicit reversible integrator with a time stem of $\Delta t = 0.48$ fs. The $g(r)$ in Fig. 2 was obtained by averaging over 0.72 ps after the initial equilibration, which takes 0.24 ps. For the tensile simulation in Fig. 4[56], the system was thermalized at 300 K in the canonical ensemble and subjected to tensile strain in the z-direction with a strain rate of $10^{10} s^{-1}$. For each applied strain, the system was simulated in the canonical ensemble for 0.48 ps. After the stress stabilized, the stress value for that strain was calculated by averaging over the last 0.29 ps.

**- Model Training**

The Allegro potential was implemented through NequIP, incorporating an E(3) symmetry equivariant neural network coded by e3nn[57] on the PyTorch framework. The model architecture, in case Allegro-FM256, consists of two layers of 64 tensor features with $l = 2$ in full O(3) symmetry. The network utilizes several multi-layer perceptrons (MLP) with specific configurations: a two-body latent MLP with dimensions [64, 128, 256] and a later latent MLP with dimensions [256, 256, 256], both employing SiLU nonlinearities. Table 1 presents the specification of each model used in this study. The embedding MLP was implemented as a linear projection, while the final edge energy MLP comprised a single hidden layer of dimension 256 without non-linearity. All MLPs were initialized according to a uniform distribution of unit variance. The training protocol employed a radial cutoff of 5.2 Å, with the loss function combining the per-atom energy, force and stress root mean square errors under 8:1:1 weight ratio. Parameter optimization was performed using the Adam optimizer. All calculations were performed using Sophia at Argonne Leadership Computing Facility.

Table 1. Hyperparameter settings and total number of trainable parameters for each Allegro-FM model.

|  | Number of tensor feature | Latent dimensions | Two_body dimensions | Edge_eng dimensions | Number of parameters |
|---|---|---|---|---|---|
| Allegro-FM64 | 16 | [64, 64, 64] | [16, 32, 64] | [64] | 44.4 K |
| Allegro-FM256 | 64 | [256,256,256] | [64,128,256] | [128] | 690 K |
| Allegro-FM512 | 128 | [512,512,512] | [128,256,512] | [128] | 2.6 M |

- Total Energy Alignment (TEA)

In this study, we extended the Total Energy Alignment (TEA) framework introduced by Shiota et al.[23] to train the Allegro neural network potential architecture, which enables massive parallel processing. TEA is a sophisticated approach that harmonizes heterogeneous quantum chemical datasets through a two-step process: Inner Core Energy Alignment (ICEA) and Atomization Energy Correction (AEC). ICEA addresses systematic energy offsets arising from different core electron treatments, while AEC scales atomization energies to account for varying computational fidelities. This methodology enables seamless integration of datasets computed under different conditions without extensive recalculations.

Our study utilized two complementary dataset proceed through the TEA method: the inorganic structure MPtrj dataset (calculated at PBE/PW level using VASP) and the OFF23 dataset of organic molecules (computed at ωB97M-D3(BJ)/def2-TZVPPD level using Psi4). The integrated dataset, publicly available through Shiota et al. 's repository[58], was preprocessed by removing structures containing noble gas elements and those with forces exceeding 0.25 Hartree to ensure stable training. This curated dataset was then used to train the Allegro architecture, which maintains E(3)-equivariant properties while enabling faster parallel computation compared to the original MACE framework.

During the training process with Allegro, we monitored both the overall loss convergence and element-wise training performance. The model achieved a root mean square error of 117 meV for per atom energies, 130 meV for forces, and 16 MPa for stress on the test dataset.

- MD Simulations of C-S-H Nanoparticle Reaction

We first have created a bulk amorphous C-S-H using CLAYFF[56], then cut out a C-S-H nanoparticle with a radius of 1.2 nm. The nanoparticle is placed in a mixture of $H_2O$ and $CO_2$ molecules with a 50:50 ratio at a density of 1 g/cc. The cubic MD system of $(4.57nm)^3$ contains 267 C, 90 Ca, 133 Si, 2577 O, and 3259 H atoms respectively. The entire system is relaxed at 300 K using NVT ensemble with 0.5 fs timestep, then subsequently heated to 1200 K in 22 ps. All simulations are performed using RXMD software [59].